\documentstyle[twoside,fleqn,espcrc2]{article}

\newcommand{\ww}{\mbox{\tiny $\wedge$}}
\newcommand{\pp}{\partial}

\newcommand{\AmS}{{\protect\the\textfont2
  A\kern-.1667em\lower.5ex\hbox{M}\kern-.125emS}}

\author{M\'aximo Ba\~nados \address{Departamento de F\'{\i}sica
Te\'orica, Universidad de Zaragoza, \\ Ciudad Universitaria, Zaragoza
50009, Spain.}}

\title{Gravitons and gauge fields in 5d Chern-Simons supergravity} 

\begin{document}

\begin{abstract}

Despite the nice geometrical properties of higher dimensional
Chern-Simons (CS) supergravity theories these actions suffer from one
major drawback, namely, their connection with the real world. After 
some quick remarks on three-dimensional gravity, we consider
five-dimensional CS supergravity and study to what extend this theory
reproduces the standard low energy description of gravitons and
gauge fields. We point out that if one deforms the CS
action by changing the value of the cosmological constant by a small
amount (thus breaking the CS symmetry), propagation around AdS becomes
non-trivial, asymptotically Schwarzschild-AdS solutions exist, and the
gauge field acquires its standard quadratic propagator.  This is the
written version of an invited Lecture delivered at the QG99 meeting,
held in Sardinia, Italy, on Sept. 1999.

\end{abstract}

\maketitle
\section{Introduction}

Chern-Simons (CS) gravity and supergravity exist in all odd 
dimensional spacetimes. The three-dimensional case, however, stands
apart. There are two reasons for this. First, the 3d CS action is
equivalent (up to some issues as the invertibility of the metric) to
the standard metric action, as opposed to the higher dimensional
CS theory which includes high order powers in the curvature tensor.
Second, the 3d theory is special due
to its great simplicity. There are many beautiful calculations that
can be done yielding interesting and surprising results, most notably,
the existence of an asymptotic conformal symmetry \cite{BH}, and its
role in understanding the 3d black hole entropy \cite{Strominger}.

Neither the conformal algebra nor its application to black hole
physics really need the CS formulation. However, this representation
provides simple derivations of some results which are otherwise
complicated. As motivation to study the higher
dimensional cases, we shall briefly review in Sec. 2 the derivation of
the Brown-Henneaux conformal algebra using the CS formulation.

Higher dimensional CS gravitational theories are far less developed
than the 3d case.  Still, it has been shown that black holes do exists
\cite{BTZ2}, and the asymptotic structure has some of the elements of
the 3d case \cite{BGH} (see, for a recent discussion,
\cite{Gegenberg-K}). In Sec. 3 we shall study whether 5d CS
supergravity has some relation with standard 5d
supergravity \cite{Chamseddine-N} (which is known to descend from
11d supergravity). We shall point out that if one
breaks the CS symmetry by changing the value of the cosmological
constant by a small amount, then solutions which are asymptotically
AdS-Schwarzschild do exist, and the gauge field entering in the
supergravity action acquires its standard propagator. 

\section{Three-dimensional gravity}

Three dimensional gravity has been analysed from many
different points of view; we refer the reader to
\cite{Carlip,Menotti,Carlip-book,Baires} for a complete
treatment of different aspects of this theory.    

In this section we will discuss a short derivation of the
Brown-Henneaux conformal algebra \cite{BH}.  This symmetry plays an
important role in 3d gravity and black hole physics \cite{Strominger}.
For a recent generalization of the following results to higher
dimensions see \cite{BCG}. 

Consider the three-dimensional anti-de Sitter line element,
\begin{equation}
{ds^2_0 \over l^2} = e^{2\rho} dw d\bar w + d\rho^2 
\label{ads0}
\end{equation}
which will be our background metric. The parameter $l$ is the AdS
radius and all coordinates are dimensionless. In (\ref{ads0}), $w$ and
$\bar w$ are coordinates on $\Re^2$ and $0\leq \rho <\infty$.  

The metric (\ref{ads0}) can be perturbed preserving the equations of
motion as
\begin{equation}
{ds^2\over l^2} = e^{2\rho} dw d\bar w + d\rho^2 + { T(w)\over k} \,
dw^2
\label{adsT}
\end{equation}
where 
\begin{equation}
k={l \over 4G}
\label{k}
\end{equation}
is the 3d (dimensionless) GR coupling constant and
$T(w)$ is an arbitrary holomorphic function of $w$. It is direct to
prove that (\ref{adsT}) is still an exact solution to the equations of
motion. The most general perturbation preserving the boundary
conditions also involves an arbitrary function of $\bar w$ \cite{BH}
(see \cite{Baires} for the explicit form of the metric), but for our
purposes here this holomorphic version will be quite useful. 

In the context of the AdS/CFT conjecture, we expect the perturbations
$T(w)$ to be related to a conformal field theory defined on the
complex plane $w$. To check this, let us act on (\ref{adsT}) with a
conformal transformation $w\rightarrow w'$, 
\begin{equation}
w=f(w')
\label{u'u}
\end{equation}
where $f$ is an arbitrary function of $w$. 

The new metric is of course still a solution of the equation of
motion, but it changes its form, 
\begin{equation}
{ds^2\over l^2} = e^{2\rho} (\pp'f) dw' d\bar w + d\rho^2 + { T(w)
\over k} (\pp' f)^2 \, dw'^2.
\label{adsT'}
\end{equation}
The discovery of Brown and Henneaux \cite{BH} is that one can put
(\ref{adsT'}) back into the original form (\ref{adsT}) via the
following redefinitions of $\rho$ and $\bar w$,
\begin{eqnarray}
e^{2\rho} &=& {e^{2\rho'} \over \pp' f} \label{rho'}\\
\bar w &=& \bar w' - {1 \over 2} e^{-2\rho'} {\pp'^2 f \over\pp' f } . 
\end{eqnarray}
In fact, inserting these new coordinates into (\ref{adsT'}) one
recovers the metric (\ref{adsT}) in the primed coordinate system with 
\begin{equation}
T'(w') = T(w) (\pp' f)^2 - {k \over 2} \{f,w'\}
\label{T'T}
\end{equation}
and where 
\begin{equation}
\{f,w\} = {\pp^3 f \over \pp f} - {3 \over 2} \left({\pp^2 f \over \pp
f}\right)^2
\end{equation}
denotes the Schwarzian derivative of $f$.  The transformation law
(\ref{T'T}) corresponds to a Virasoro operator with central charge 
\begin{equation}
c= 6k = {3l \over 2G}.
\label{c}
\end{equation}
See \cite{CFT} for an extensive recent exposition on two dimensional
conformal field theory. 

Note that when applying the conformal transformation (\ref{u'u}) one
also needs to change the radial coordinate. This is the Brown-Henneaux
version of the IR-UV correspondence discussed in \cite{Susskind-W}.   

This derivation of (half of) the Brown-Henneaux symmetry is
particularly interesting for two reasons. First, it deals with a
finite conformal transformation showing clearly that anti-de Sitter
space knows very well the Schwarzian derivative. Second, it acts in a
natural way on the space of solutions of the theory mapping solutions
into solutions.

Now, there is a very important point that we have omitted so far. We
have found the transformation properties of $T$ via a simple 
transformation of coordinates. Why should we expect this coordinate
redefinition to carry relevant information?  The reason is that
(\ref{u'u}) does not go to the identity at the (conformal) boundary
$\rho \rightarrow \infty$ of anti-de Sitter space. Indeed, $f$ does
not depend on $\rho$ at all. It is well known that those gauge
transformations that do not go fast enough to zero at the boundary may
represent non-trivial degrees of freedom. 

A powerful way to understand this point is the Regge-Teitelboim
approach \cite{Regge-T}. The idea is that a proper gauge
transformation is generated by a constraint and act on physical states
by annihilation $G|\phi\rangle=0$. The state is thus invariant.
However, those gauge transformations whose parameters do not vanish
fast enough at the boundary are not generated by constraints and it is
then inconsistent to assume that they leave physical states invariant.
The transformation $w\rightarrow w'$ is an example of this class of
symmetries. A corollary of this result is that  two metrics of the
form (\ref{adsT}) with different values of $T$ represent physically
different solutions to the equations of motion.

This issue, applied to 3d gravity, has been analysed in detail in
\cite{BH} where it was shown explicitly that the conformal map
$w\rightarrow w'$ is not generated by a constraint. We shall now
make the transition to the CS formulation and review a simple
derivation of this fact.  The rest of this section is well-known
material and we include it here for completeness. We shall follow
mainly \cite{B} and \cite{CHvD}, and assume some familiarity with the
Chern-Simons formulation of three dimensional gravity
\cite{Achucarro-T,Witten88}. 

We shall first discuss the appearance of an affine Kac-Moody algebra
(at the boundary) in any CS theory with a group ${\cal G}$. 

The canonical generator $G(\lambda)$ associated to a gauge
transformation, $\delta A^a= D\lambda^a = [A^a,G(\lambda)]$, in any
three-dimensional Chern-Simons theory is given by
\cite{Balachandran,B},
\begin{equation}
G(\lambda) = {k \over 4\pi}\int_\Sigma \lambda_a F^a - {k \over
4\pi} \int_{\pp \Sigma} \lambda_a A^a .
\label{G}
\end{equation}
Here $\Sigma$ is a two dimensional spatial section with boundary
$\pp\Sigma$, $F$ is the 2-form curvature and $A$ the gauge field.
The equations of motion of CS theory are $F=0$ showing that the bulk
part of (\ref{G}) is indeed a constraint, however, the boundary piece
does not vanish on shell. 

If $\lambda$ is zero and $A$ is finite at $\pp\Sigma$ then the
boundary term in (\ref{G}) is zero. This corresponds to a proper gauge
transformation which does not change physical states. But the
action is also invariant under transformations with non-zero
values of $\lambda$ at the boundary\footnote{Provided some boundary
conditions are obeyed. For example, one can impose the chiral boundary
condition $A_{\bar w}=0$; the residual group contain all parameters
which at the boundary only depend on $w$, but are not necessarily zero
there.}. The canonical formalism is powerful enough to generate gauge
transformation whose parameters do not vanish at the boundary provided
the charge 
\begin{equation}
J(\lambda) = -{k \over 4\pi} \int_{\pp \Sigma} \lambda_a A^a
\end{equation}
is finite. In this case, the corresponding transformation is not
``gauge" because it is not generated by a constraint.    

The charge $J$ is not a gauge invariant quantity
because $A$ transforms inhomogeneously under gauge transformations. We
act on $J$ with a gauge transformation $\delta_\rho A^a=D\rho^a =
d\rho^a + [A,\rho]^a$ obtaining, 
\begin{eqnarray} 
\delta_\rho J(\lambda) &=& -{k \over 4\pi}\int_{\pp \Sigma}
([\lambda,\rho]_a A^a +  \lambda_ad\rho^a)  \nonumber\\
&=& J([\rho,\lambda]) - {k \over 4\pi}\int_{\pp
\Sigma}\lambda_ad\rho^a
\label{KM}
\end{eqnarray}
On the other hand, as discussed in
\cite{Regge-T,Benguria-CT,Henneaux-T,BH}, the variation of the
charge under an allowed transformation is itself given by the
commutator of two charges,  
\begin{equation}
\delta_\rho J(\lambda) = [J(\rho),J(\lambda)].
\label{pbs}
\end{equation}
Eqns. (\ref{KM}) and (\ref{pbs}) show that the charge $J$
satisfies an affine Kac-Moody algebra with central charge $k/2$. The
appearance of an affine algebra at the boundary was first pointed out
in \cite{Witten89} in relation to CS theory and the Jones polynomial. 
 
This derivation of the affine algebra is valid for any group ${\cal
G}$. To apply these results to 3d gravity, we consider the particular
case ${\cal G}=SL(2,\Re) \times SL(2,\Re)$ and identify the CS
coupling $k$ with the gravitational coupling (\ref{k}). 

The next step is to identify the origin of the Brown-Henneaux Virasoro
symmetry in this discussion. Given an affine algebra, there is a
natural Virasoro operator associated to it via the Sugawara
construction. The central charge in the Sugawara construction is
however quantum mechanical and thus it cannot be identified with
(\ref{c}). It was shown in \cite{B} that a twisted Sugawara operator
yields the correct Brown-Henneaux algebra. The twisting is related to
the radial diffeomorphism (\ref{rho'}) necessary to bring the metric
(\ref{adsT'}) back to its original form. This derivation is however
not complete because it does not fix the value of the central charge.
Indeed, from the purely CS point of view, the coefficient in the
twisting can be arbitrary and thus the central charge is arbitrary. A
full derivation of the conformal algebra in the CS formulation was
presented in \cite{CHvD}. The idea is to use the reduction of affine
SL(2,R)$_k$ to Virasoro discussed in \cite{Polyakov}. This follows by
looking at the metric (\ref{adsT}) and choosing boundary conditions
for the gauge field such that the associated metric has the asymptotic
behaviour (\ref{adsT}). These boundary conditions coincide exactly
with the reduction conditions considered in \cite{Polyakov}. The
Virasoro symmetry then follows in a direct way with the correct value
for the central charge $c=6k$.            

The details of this derivation can be found in \cite{CHvD}. See also
\cite{BBO,BO} for a comparison with \cite{B}.

\section{Five dimensional Chern-Simons supergravity}

\subsection{The action}

Contrary to the three-dimensional case, the 5d Chern-Simons (CS) 
supergravity action is quite different from the standard theory. This
theory was first studied in \cite{Chamseddine}; generalizations to
other dimensions have been studied in \cite{TZ}.      

Despite the nice geometrical properties of the CS construction, this
theory has one major drawback, namely, its connection with the real
world.  For this theory to be ``phenomenologically" reasonable, i.e.,
related to some compactification of 11d supergravity or $M$ theory,
one should be able to match its field content with the usual modes of
5d supergravity, namely, the graviton $h_{\mu\nu}$, a graviphoton
$A_\mu$, plus some spinor fields \cite{Chamseddine-N}. While the 5d CS
action does contain fields with those tensorial properties, their
propagators are not the standard ones.

We start by briefly reviewing the construction of the action
\cite{Chamseddine}. 

For those familiar with the CS construction and super Lie algebras, it
should be enough to say that 5d CS supergravity is simply the integral
of a CS density for the supergroup $SU(2,2|N)$.  This supergroup is
the smallest super extension of $SO(4,2)$, which is the group
associated to the pure gravitational part. It is, however, of some
pedagogical value to start from the purely gravitational piece and
motivate the appearance of $U(2,2)$ at the bosonic level. We refer the
reader to \cite{Chamseddine,BTZ2,TZ} for more details. In the
following, we assume  some familiarity with the bosonic Chern-Simons
construction.    

The purely gravitational piece contains the $SO(4,2)$ gauge field
\begin{equation}
W_G = e^a P_a + {1 \over 2} w^{ab} J_{ab}
\end{equation}
where $P_a$ and $J_{ab}$ are the
$SO(4,2)$ generators, and $e^a$ and $w^{ab}$ are the veilbein and spin
connection respectively. The subscript $G$ refers to ``Gravitational".
As usual when studying supergravity, we
consider the particular representation of $SO(4,2)$ obtained via
Dirac matrices,
\begin{equation}
P_a = {1 \over 2} \gamma_a, \ \ \  J_{ab} = {1 \over 2}
\gamma_{ab}, \ \ \  \{\gamma_a,\gamma_b\} = 2\eta_{ab}.
\end{equation}
Here $\gamma_{ab}$ is the anti-symmetrised product of Dirac matrices
normalized by $\gamma_{ab}=\gamma_a \gamma_b$ for $a\neq b$.  We use
the representation on which, 
\begin{equation}
\gamma_0=i\, \mbox{diag}(1,1,-1,-1),
\label{g0}
\end{equation}
and the signature is $\eta_{ab} = \mbox{diag}(-1,1,1,1,1)$. The matrix
$\gamma_0$ is anti-Hermitian while the others $\gamma_a$, $a\neq
0$, are Hermitian.  

Using the Dirac matrix algebra it is direct to prove that
\begin{equation}
(\gamma_a \gamma_0)^\dagger = \gamma_a \gamma_0, \ \ \ \ \  
(\gamma_{ab} \gamma_0)^\dagger = \gamma_{ab} \gamma_0. 
\end{equation}
Thus, in this representation, the gravitational Chern-Simons field
$W_G=(1/2) (e^a \gamma_a + (1/2) w^{ab} \gamma_{ab})$ belongs to the
Lie algebra of $SU(2,2)$ since
\begin{equation}
(W\gamma_0)^\dagger = W\gamma_0,
\label{her}
\end{equation}
and Tr$\,W=0$. What we have done is to ``discover" that the Lie
algebras
of $SO(4,2)$ and $SU(2,2)$ are isomorphic.  We note that although in
four dimensions the Chern-Simons action does not exist, a similar
analysis at the level of the Lie algebras, plus the existence of
Majorana spinors, yields the relation between $SO(3,2)$ and $Sp(4)$.  

The group $SU(2,2)$ admits a supersymmetric extension.  The first
step is to add an extra bosonic piece $A$ which induces a
non-zero trace but preserves (\ref{her}). We consider the extended
$U(2,2)$ bosonic gauge field,  
\begin{equation}
W = iA\, I + {1 \over 2} e^a \gamma_a + {1 \over 4} w^{ab}
\gamma_{ab}
\end{equation}
where $A$ is an Abelian gauge field and $I$ is the identity in the
space of Dirac matrices. 

A motivation for the field $A$ can be found in \cite{BTrZ} where it
was shown that, in the zero cosmological constant case, this field is
enough to achieve supersymmetry.  If the cosmological constant is not
zero, the structure of the supergroup in five dimensions induces 
extra bosonic $SU(N)$ gauge fields which however do not couple to the
gravitational variables.  

The full 5d CS supergravity action has the form \cite{Chamseddine,TZ} 
\begin{eqnarray}
I_{SU(2,2|N)} &=& I_{U(2,2)}+ I_{SU(N)}  +
\mbox{fermionic} \nonumber\\
&& \mbox{pieces + interaction terms}. 
\end{eqnarray}

We shall consider here the situation on which all fermionic and
$SU(N)$ gauge fields are zero. The action reduces to, 
\begin{equation}
I_{U(2,2)} = k( I_G + I_A + I_I)
\label{CS}
\end{equation}
where
\begin{eqnarray}
I_G &=& \int \epsilon_{abcde} \left({1 \over l} R^{ab}\ww R^{cd}\ww
e^e + \label{CSG}\right. \\ && \left.  {2 \over 3l^3 } R^{ab}\ww e^c
\ww e^d \ww e^e  + {1 \over 5 l^5 } e^a\ww e^b\ww e^c\ww e^d\ww e^e
\right) \nonumber 
\end{eqnarray}
is the purely gravitational piece, $I_A= \int A FF$ is the Abelian CS
action, and $I_I$ the interaction term,
\begin{equation}
I_I = \int \left[{1 \over 2}R^{ab}\ww R_{ab}-d(e_a T^a)\right] \ww A.
\label{CSA}
\end{equation}
From the point of view of CS theory this last term describes the
interaction between the gravitational variables and the Abelian gauge
field $A$. However, as we shall see soon, the kinetic term for $A$ is
actually contained in (\ref{CSA}).      

$k$ is a dimensionless number representing the CS coupling, and
$l$ is the AdS radius. The effective five dimensional Newton
constant and cosmological constant are (up to numerical factors), 
\begin{equation}
{1\over G} = {k \over l^3}, \ \ \ \ \ \ \ \ \Lambda = {1 \over l^2}. 
\end{equation}
The gravitational part (\ref{CSG}) can be written schematically as,
\begin{equation} 
I_G = {1 \over G} \int \sqrt{-g}\left( R + {1 \over l^2} + l^2 R\cdot
R \right) 
\label{IG}
\end{equation}
where $R\cdot R$ denotes the Gauss-Bonnet term,
\begin{equation}
R\cdot R = R^2-4R^{\mu\nu}R_{\mu\nu} +
R^{\mu\nu\lambda\rho}R_{\mu\nu\lambda\rho},
\end{equation}
which in five dimensions is not a total derivative. The first two
terms in (\ref{IG}) are of course the usual Einstein-Hilbert and
cosmological constant pieces. Our main goal in this note is to analyse
whether the quadratic interaction can be discarded or not when
expanding around some background.  

The motivation to look for phases on which the piece $R\cdot R$ does
not contribute is the existence of another 5d supergravity action
whose bosonic part has the form \cite{Chamseddine-N},  
\begin{equation}
I_{Sugra} = \int [\sqrt{-g} (R + F^{\mu\nu} F_{\mu\nu}) + F\ww F\ww
A]. 
\label{Sugra}
\end{equation}
(The exact coefficients can be found in \cite{Chamseddine-N}.) This
action is known to descend form eleven dimensional supergravity via
compactification\footnote{It is amusing to note that there also
exists a CS action for supergravity in eleven dimensions \cite{TZ}
constructed with the group $OSp(32|1)$ which, by the way, is believed
to be the
symmetry group of M theory.  Somehow, CS theories incorporate in a
very natural way the kinematics of M theory but not its dynamics. See
\cite{Horava} for a discussion on 11d CS supergravity and M-Theory.}. 

Besides the obvious differences between (\ref{CS}) and (\ref{Sugra}),
they do have one thing in common, namely, they both contain the
graviton $e^a_\mu$ and an Abelian gauge field $A_\mu$ with a CS
term.  Actually, the name graviton for the field $e^a_\mu$ is somehow
premature because, due to the quadratic Gauss-Bonnet interactions, the
spin connection is an independent field in this theory \cite{BGH}.  In
other words, it is not true that the variation of $I_{G}$ with respect
to $w^{ab}$ yields an algebraic equation. That equation is in fact
differential and carry its own dynamics. (Although $T^a=0$ is still a
solution, at least in vacuum space.)   

If we expect (\ref{CS}) and (\ref{Sugra}) to be related, we will need
to resort to some mechanism that can enable us to discard the
quadratic interactions in (\ref{CS}). However, even if we could argue
that the higher order interactions are negligible in some limit, we
will then face the problem that the kinetic piece for $A$ in
(\ref{CS}) is not the standard one. In fact, besides the Abelian CS
interaction, $A$ does not seem to have a kinetic piece at all! ~ We
shall see that these two problems are closely related and that if we
can solve the first, the second will be solved automatically.

\subsection{Backgrounds}

Let us first study whether one could neglect the quadratic interaction
terms in (\ref{IG}).  We set for the moment $A=0$ and deal only with
(\ref{CSG}), or in its metric form (\ref{IG}). As we shall see, this
may not be the right way to analyse this problem, but it is a good
starting point.  

We shall study solutions to the equations of motion with zero torsion.
If the quadratic interaction was ``small" in some limit, one would
expect, for example, to find solutions which asymptotically behave
like AdS-Schwarzschild. A situation like this occurs in Born-Infeld
electrodynamics whose spherically symmetric solutions approach the
standard Coulomb potential. In that case one can argue that at low
energies the Born-Infeld action reduces to the standard one. This does
not happen
in CS gravity.

Solutions with spherical symmetry for the action (\ref{CS}), with
$A=0$, have been studied in \cite{BTZ2}. The metric has the form,
\begin{equation}
ds^2 = -f(r) dt^2 + {dr^2 \over f(r)} + r^2 d\Omega_3
\label{sph-sol}
\end{equation}
with
\begin{equation}
f(r) = -C + {r^2\over l^2} . 
\label{BTZ2}
\end{equation}

The parameter $C$ is a constant of integration, related to the CS-ADM
energy of the solution (see below). If $C=-1$, this solution 
reduces to AdS space. For $C \neq -1$ there is a curvature singularity
at $r=0$, however, if $C>0$ this singularity is protected by an event
horizon.  The metric (\ref{BTZ2}) is a natural generalization to
higher dimensions of the three dimensional black hole \cite{BTZ}. We
note in particular that the singularity is smoother that the
Schwarzschild one. The curvature near $r=0$ blows up as $1/r^2$ as
opposed to Schwarzschild in five dimensions which goes like $1/r^4$.
Furthermore, at the equator $\theta_1=\theta_2=\pi/2$, the metric
(\ref{BTZ2}) is exactly a three-dimensional black hole \cite{BTZ},
i.e., 3d anti-de Sitter space with identifications \cite{BHTZ}.   
See \cite{Ilha,Myers,Kiefer,Cai} for other aspects of higher
dimensional CS black holes.  

The Schwarzschild spacetime in five dimensions has the form
(\ref{sph-sol}) with $f_{Sch}(r) = 1 + r^2/l^2 - 2M/r^2$. It is clear
that $f(r)$ given in (\ref{BTZ2}) does not approach $f_{Sch}(r)$ at
large values of $r$. We conclude that asymptotically one cannot
neglect the quadratic term. At least not when expanding around the
anti-de Sitter background.

The AdS background seems to be the most natural choice for CS 
theory. However, as stressed in \cite{BGH,Paniak}, it is degenerate
because there is no propagation around it.  Indeed, the CS equations
of motion are,
\begin{equation}
\epsilon_{abcde} (R^{ab} + {1 \over l^2} e^a  e^b
)(R^{cd} + {1 \over l^2} e^c  e^d ) = 0,
\label{ECS}
\end{equation}
and the AdS background satisfies $R^{ab} + {1 \over l^2} e^a  e^b=0$. 
Certainly there can be no linearized perturbations around this
background.  

It has been shown in \cite{BGH} that non-degenerate backgrounds for
the equations (\ref{ECS}) do exists and some examples were displayed
explicitly in that reference. However, those examples were rather
unphysical and did not have a geometrical motivation.      

In this contribution, we will change the point of view. Instead of
changing the background, we will change the action. The idea is to
deform the equations of motion a little bit such that propagation
around AdS becomes non-trivial. In particular, we will modify the
value of the cosmological constant. Hopefully, this deformation can be
achieved by turning on some of the other fields in the theory and
letting them to have a non-zero vacuum expectation value. Work in
this direction is in progress \cite{MB}. Here we shall motivate this
procedure and display its consequences.    

We deform the CS equations (\ref{ECS}) by adding a little bit of
cosmological constant of magnitude $\varepsilon^2$,    
\begin{equation}
\epsilon_{abcde}(R^{ab} + {1+\varepsilon \over l^2} e^a
e^b)(R^{cd} + {1 - \varepsilon \over l^2} e^c  e^d) = 0
\label{dE}
\end{equation}
where $\varepsilon$ is the deformation parameter. In the deformed
theory there are two ``natural" backgrounds with different AdS radii.
We will see that the corresponding spectra are also different.
Consider first the ``black hole" branch whose background is    
\begin{equation}
R^{ab} + {1 - \varepsilon \over l^2} e^a \ww e^b=0.
\label{back}
\end{equation}
Expanding the equations of motion to first order in this background we
find 
\begin{equation}
2 \varepsilon \, \epsilon_{abcde} \left[\delta R^{ab} +
{2(1-\varepsilon)\over l^2} e^a \delta e^b\right] e^c e^d =0 .
\label{dE'}
\end{equation}
This equation coincides exactly with the linearised perturbations
following from the standard 5d Einstein-Hilbert action with a
cosmological constant. An important point, however, is the presence of
the deformation parameter multiplying (\ref{dE'}). The effective
Einstein theory has a deformed Newton constant $G_\varepsilon$ and AdS
radius $l_\varepsilon$ given by
\begin{equation}
G_\varepsilon = {G \over \varepsilon}, \ \ \ \ \ \ \ \ {1 \over
l^2_\varepsilon} = {(1-\varepsilon) \over l^2} .
\label{defp}
\end{equation}
In particular, we cannot let $\varepsilon\rightarrow 0$ without
loosing the contact with standard Einstein equations.  We have thus
shown that the dynamics of the deformed CS theory expanded around
(\ref{back}) is described by the action,
\begin{equation}
I \sim {1 \over G_\varepsilon} \int  \sqrt{-g} \left( R + {2 \over
l_\varepsilon^2} \right).
\label{IGR}
\end{equation}

Let us now consider solutions with spherical symmetry in the deformed
theory.  Exact solutions for the equations (\ref{dE}) with 
spherical symmetry and $\varepsilon$ not necessarily small are known
\cite{Boulware-D}. They have the form (\ref{sph-sol}) with,
\begin{equation}
f_{\varepsilon}(r) = 1 + {r^2 \over l^2} - 
\sqrt{ { r^4 \over l^4} \,\varepsilon^2 + 4 M G }.   
\label{df}
\end{equation}
If $\varepsilon=0$, we recover the previous solution with
$C=2\sqrt{MG}$.  Note that we have made a choice in the sign in
front of the square root such that, for $M=0$, we recover the
background (\ref{back}). This choice also ensures the existance of the
horizon and this is why we call it the ``black hole" branch. Had we
chosen the other background, with $l^2_\varepsilon =
l^2/(1+\varepsilon)$, the corresponding solutions would be
``particles" \footnote{This interpretation follows
from the fact that singularity at $r=0$ in CS theories is smoother
than that of standard GR. This effect comes from the quadratic
interaction which is dominant near the origin.}.  The two backgrounds
of (\ref{dE}) then lead to two branches of the theory; one containing
black holes and the other containing particles.  

The solution (\ref{df}) is indeed asymptotically Schwarzschild since
for large values of $r$, $f_{\varepsilon}(r)$ behaves as 
\begin{equation}
f_{\varepsilon}(r) = 1 + {r^2 \over l_\varepsilon^2} - {2m
G_\varepsilon \over r^2} + {\cal O}({1 \over r^6}).
\label{Sch}
\end{equation} 
where $G_\varepsilon$ and $l_\varepsilon$ are given in (\ref{defp}),
and $m = Ml^2$ is the ADM mass. Thus, asymptotically,
the solutions to (\ref{dE}) are indeed solutions to (\ref{IGR}).

Before leaving this section we would like to analyse in some more 
detail the definition of the ADM energy in CS theories and in
particular its relation with the integration constant $M$.  

A quick way to analyse this problem is to perform a minisuperspace
reduction \cite{BTZ2}. Consider the space of metrics of the form
(\ref{sph-sol}) for any function $f$. Let $ f(r) = 1 + h(r)$.
Plugging
the metric into the equations (\ref{dE}) it is direct to see that the
only non-trivial equation is
\begin{equation}
{\cal H} = -{1 \over 4G f^{1/2}}{d \over dr} \left[ \left( {r^2\over
l^2} -
h\right)^2 - \varepsilon^2 { r^4 \over l^4 } \right] =0.
\label{H}
\end{equation}
This equation is of course trivially integrated and leads to
(\ref{df}).  Eq. (\ref{H}) corresponds to the variation of the CS
action with respect to the lapse function $N = (-g_{00})^{1/2}$.
Indeed, ${\cal H}$ is the Hamiltonian constraint which enters in the
action as $H = \int N^\perp {\cal H}  + E$ where $E$, the ADM energy,
is a boundary term added to make $H$ differentiable \cite{Regge-T}.
For this class of metrics, $E$ is given by 
\begin{equation}
E = {1 \over 4G} \left[ \left( {r^2\over l^2} -
h\right)^2 - \varepsilon^2 { r^4 \over l^4 } \right]_{r\rightarrow
\infty}.
\label{E}
\end{equation}
Here we have adjusted a constant such that $E=0$ on the AdS
background $h_{(AdS)} = (1-\varepsilon) r^2/l^2$.  

The value of $E$ on the solution (\ref{df}) is $E=M$ showing that the
parameter $M$ is indeed the CS-ADM energy.  

It is instructive to look separately at each piece in (\ref{H}) and
(\ref{E}). Consider $(r^2-h)^2 = r^4 -2 hr^2 + h^2$.  The piece $r^4$
is the contribution to the cosmological constant to the Hamiltonian
\footnote{This piece does not in principle contribute to the energy,
however, since we fix $E_{AdS}=0$, the substraction brings this
piece back into $E$.}. The piece $r^2 h$ is the contribution from the
Einstein Hilbert term, giving a finite contribution for perturbations
of the form $1/r^2$. Finally, the piece $h^2$ is the contribution to
the Hamiltonian coming from the Gauss-Bonnet quadratic interaction.

\subsection{The gauge field propagator}  

We have shown in the last paragraph that the quadratic interactions in
the CS Lagrangian, in general, cannot be neglected. However, we have
also noticed that a small deformation in the value of the
cosmological constant did produce an asymptotically Schwarzschild
solution and more generally, a set of perturbations described by
standard GR with the ``renormalised" couplings (\ref{defp}).  This
means that asymptotically, the deformed action is controlled by the
Hilbert piece and, in that case, one can neglect the 
quadratic piece, provided one renormalises Newton constant
appropriately.  Deforming the cosmological term means that the CS
structure is lost, and in particular, its symmetries are broken.
(Actually, only the AdS translational symmetry is broken;
Diffeomorphisms and Lorentz rotations are not affected.) We loose the
nice geometrical properties but we gain a phenomenological space of
solutions, namely, asymptotically Schwarzschild spacetimes.  

As further motivation to consider breaking the Chern-Simons symmetry,
we shall now show that on this phase, the Abelian gauge field $A$
which enters in the CS action (\ref{CS}) in a rather strange form,
becomes the usual $U(1)$ electromagnetic field.   

We shall prove that if we discard the terms with more that two
derivatives in (\ref{CSA}), in particular the Gauss-Bonnet
interaction, then the term $d(e_a T^e) \ww A$ appearing
in (\ref{CSA}) is exactly equivalent to the quadratic propagator
$\sqrt{-g} F^{\mu\nu}F_{\mu\nu}$. 

The mechanism leading to the appearance of $F^{\mu\nu}F_{\mu\nu}$
through a first order action is actually a general feature of p-forms
in $p+4$ dimensions and can be described aside of the supergravity
action. The simplest example occurs in four dimensions with a
0-form (scalar) coupled to gravity via the action, 
\begin{equation}
I[e,w,\phi] = \int \epsilon_{abcd} R^{ab}\ww e^c \ww e^d + 
d(e_a\ww T^a) \phi,
\label{Iphi}
\end{equation}
where $T^a = De^a$ is the torsion tensor\footnote{The combination
$d(e_a T^a)=T^a T_a + R_{ab} e^a\ww e^b$ is sometimes called the
Nieh-Yan invariant.  Its relation to chiral anomalies in four
dimensions has been discussed in \cite{CZ}.}. The scalar
field $\phi$ appears in (\ref{Iphi}) with no derivatives. We shall now
show that (\ref{Iphi}) is equivalent to the standard  $\int \sqrt{g}
(R + (\pp_\mu \phi)^2)$ action.   

Since $e_a T^a$ is linear in the spin connection, its variation is
independent of it.  The equation of motion that follows from varying
(\ref{Iphi}) with respect to $w^{ab}_\mu$ is an algebraic
equation 
\begin{equation}
e_{a\mu} T^a_{\nu\lambda} = \epsilon_{\mu\nu\lambda\rho} \pp^\rho \phi 
\label{T}
\end{equation}
which can be solved for $w^{ab}_\mu$. The spin connection separates in
the torsion-free part plus a torsion contribution 
\begin{equation}
w^{ab}_\mu =w^{ab}_\mu(e) + e^{a\nu}e^{b\lambda}
\epsilon_{\mu\nu\lambda\rho} \pp^\rho \phi
\end{equation}
Here $w(e)$ is the standard formula that follows from $T^a=0$. 
We now eliminate $w$ by replacing its on-shell value back in the
action.  An straightforward calculation shows that one indeed obtains
the standard scalar field action coupled to gravity. A quick way to
convince ourselves that this is true is by considering the
$\phi-$equation of motion following from (\ref{Iphi}), 
\begin{equation}
d(e_a T^a) =0. 
\label{phi-eq}
\end{equation}
Replacing (\ref{T}) in this equation one derives the correct scalar
field equation $\pp_\mu(\sqrt{g} g^{\mu\nu} \pp_\mu \phi)=0$.   

The key property of (\ref{Iphi}) is that it does not need the metric
tensor explicitly to achieve diffeomorphism invariance. This is a
characteristic feature of Chern-Simons Lagrangians.  The action
(\ref{Iphi}) can be regarded as a first order formulation of an 
scalar field couple to gravity. The metric is not explicitly needed
(although present through the tetrad field) and in that sense
(\ref{Iphi}) is topological.   

This construction can be generalised to higher dimensional situations
involving p-forms. The next case is precisely the action (\ref{CSA})
on which we neglect the quadratic pieces in the Riemann curvature.    
The Chern- Simons interaction $FFA$ does not depend on the spin
connection. The variation with respect to the spin connection then
yields 
\begin{equation}
e_{a\mu} T^a_{\nu\lambda}= \epsilon_{\mu\nu\lambda\rho\sigma}
F^{\rho\sigma}
\end{equation}
Eliminating the spin connection one recovers the standard coupling
$\int \sqrt{g} (R + F^{\mu\nu} F_{\mu\nu})$ (plus the $FFA$
interaction). Again, a quick way to check this assertion is 
by considering the variation of the action with respect to $A_\mu$
which yields again $d(e_a \ww T^a) =0$. Since the torsion equation
determines $e_a \ww T^a = ~^*F$, the above equation becomes $d^*F=0$,
as desired.  This mechanism leading to the standard propagator for the
gauge field has already appeared in the group manifold approach to
five-dimensional supergravity \cite{Castellani}.   

The coupling of a $p-$form in $d=4+p$ dimensions can be treated in a
unified way in this fashion. The field equation will always be $d(e_a
\ww T^a)=0$ which, supplemented with the torsion equation $e_a \ww T^a
= ~^*F_{p+1}$, yields $d^*F_{p+1}=0$ as desired. We end by mentioning
that this mechanism is well-known in the string theory literature. The
field $H_{\mu\nu\lambda}$ sometimes called the ``torsion" receives
this name precisely because it can be absorbed into the spin
connection as we have described here.

\section{Conclusions and open problems} 

We have shown in this contribution that deforming five dimensional CS
supergravity leads to a new theory which is closer to standard
gravitational theories. There are many open problems which we have not
touched here. Most importantly, we have not really proved that the
deformed theory reduces to the standard supergravity theory, but only
checked that the bosonic degrees of freedom have the usual dynamics.
It would be very interesting to see whether the full CS supergravity
action admits a deformation such that, for some value of the
parameter, it reproduces the standard supergravity action. Note that
supersymmetry fixes all relative coefficients and thus this would 
uncover a curious relation between CS and standard supergravity.\\
\\ 

It is a great pleasure to thank the organizers of QG99 for the
invitation to participate in the last Quantum Gravity meeting of the
millennium. Besides the great environment and fruitful discussions
that I enjoyed during the conference, I also had the undeserved
privilege to win a bottle of good Sardinian wine for which I am most
grateful. I would also like to thank Manuel Asorey, Andrew Chamblin,
Fernando Falceto, Gary Gibbons, Marc Henneaux, Ricardo Troncoso and
Matt Visser for useful conversations.  Financial support from CICYT
(Spain) grant AEN-97-1680, and the Spanish postdoctoral program of
Ministerio de Educaci\'on y Cultura is also acknowledge.

\end{document}